\documentclass[%
 reprint,
superscriptaddress,
 amsmath,amssymb,
 aps,
]{revtex4-2}

\usepackage{graphicx}
\usepackage{float}
\usepackage{dcolumn}
\usepackage{bm}
\usepackage{physics}
\usepackage{xcolor}
\usepackage{siunitx}
\usepackage{soul}
\usepackage{amsmath}

\definecolor{gio}{rgb}{0.82, 0.34,0.80}
\definecolor{NiCCo}{rgb}{1,0,0.5}
\definecolor{luca}{rgb}{0,0,1}
\definecolor{nic}{rgb}{0.1,0.9,0.85}

\definecolor{giu}{rgb}{1,0.7,0.7}

\graphicspath{{./figure/}}

\renewcommand{\dd}{\mathrm{d}}

\usepackage[colorlinks=true, citecolor = blue, linkcolor = blue, urlcolor=blue, pdfborder={0 0 0}]{hyperref}

\usepackage[normalem]{ulem}

\begin{document}

\preprint{APS/123-QED}

\title{Single-fluid model for rotating annular supersolids and its experimental implications}
\author{N. Preti}
\affiliation{Dipartimento di Fisica e Astronomia, Università degli Studi di Firenze, Via G. Sansone 1, 50019 Sesto Fiorentino, Italy}
\affiliation{European Laboratory for Nonlinear Spectroscopy (LENS), Via Nello Carrara 1, 50019 Sesto Fiorentino, Italy}
\affiliation{Istituto Nazionale di Ottica, Consiglio Nazionale delle Ricerche (INO-CNR), Via Moruzzi 1, 56124 Pisa, Italy}
\author{N. Antolini}
\affiliation{European Laboratory for Nonlinear Spectroscopy (LENS), Via Nello Carrara 1, 50019 Sesto Fiorentino, Italy}
\affiliation{Istituto Nazionale di Ottica, Consiglio Nazionale delle Ricerche (INO-CNR), Via Moruzzi 1, 56124 Pisa, Italy}
\author{C. Drevon}
\affiliation{European Laboratory for Nonlinear Spectroscopy (LENS), Via Nello Carrara 1, 50019 Sesto Fiorentino, Italy}
\affiliation{Istituto Nazionale di Ottica, Consiglio Nazionale delle Ricerche (INO-CNR), Via Moruzzi 1, 56124 Pisa, Italy}
\author{P. Lombardi}
\affiliation{Istituto Nazionale di Ottica, Consiglio Nazionale delle Ricerche (INO-CNR), Largo Enrico Fermi 2, 50125 Firenze, Italy}
\affiliation{European Laboratory for Nonlinear Spectroscopy (LENS), Via Nello Carrara 1, 50019 Sesto Fiorentino, Italy}
\author{A. Fioretti}
\affiliation{Istituto Nazionale di Ottica, Consiglio Nazionale delle Ricerche (INO-CNR), Via Moruzzi 1, 56124 Pisa, Italy}
\author{C. Gabbanini}
\affiliation{Istituto Nazionale di Ottica, Consiglio Nazionale delle Ricerche (INO-CNR), Via Moruzzi 1, 56124 Pisa, Italy}
\author{G. Ferioli}
\affiliation{Dipartimento di Fisica e Astronomia, Università degli Studi di Firenze, Via G. Sansone 1, 50019 Sesto Fiorentino, Italy}
\affiliation{European Laboratory for Nonlinear Spectroscopy (LENS), Via Nello Carrara 1, 50019 Sesto Fiorentino, Italy}
\author{G. Modugno}
\affiliation{Dipartimento di Fisica e Astronomia, Università degli Studi di Firenze, Via G. Sansone 1, 50019 Sesto Fiorentino, Italy}
\affiliation{European Laboratory for Nonlinear Spectroscopy (LENS), Via Nello Carrara 1, 50019 Sesto Fiorentino, Italy}
\affiliation{Istituto Nazionale di Ottica, Consiglio Nazionale delle Ricerche (INO-CNR), Via Moruzzi 1, 56124 Pisa, Italy}
\author{G. Biagioni}
\email{giulio.biagioni@institutoptique.fr}
\affiliation{Dipartimento di Fisica e Astronomia, Università degli Studi di Firenze, Via G. Sansone 1, 50019 Sesto Fiorentino, Italy}
\affiliation{European Laboratory for Nonlinear Spectroscopy (LENS), Via Nello Carrara 1, 50019 Sesto Fiorentino, Italy}
\affiliation{Istituto Nazionale di Ottica, Consiglio Nazionale delle Ricerche (INO-CNR), Via Moruzzi 1, 56124 Pisa, Italy}
\affiliation{Universit\'e Paris-Saclay, Institut d'Optique Graduate School, CNRS, Laboratoire Charles Fabry, 91127, Palaiseau, France}

\begin{abstract}
The famous two-fluid model of finite-temperature superfluids has been recently extended to describe the mixed classical-superfluid dynamics of the newly discovered supersolid phase of matter. We show that for rigidly rotating supersolids one can derive a more appropriate single-fluid model, in which the seemingly classical and superfluid contributions to the motion emerge from a spatially varying phase of the global wavefunction. That allows to design experimental protocols to excite and detect the peculiar rotation dynamics of annular supersolids, including partially quantized supercurrents, in which each atom brings less than $\hbar$ unit of angular momentum. Our results are valid for a more general class of density-modulated superfluids.           
\end{abstract}

\maketitle

\begin{figure}
    \centering
    \includegraphics[width=1\columnwidth]{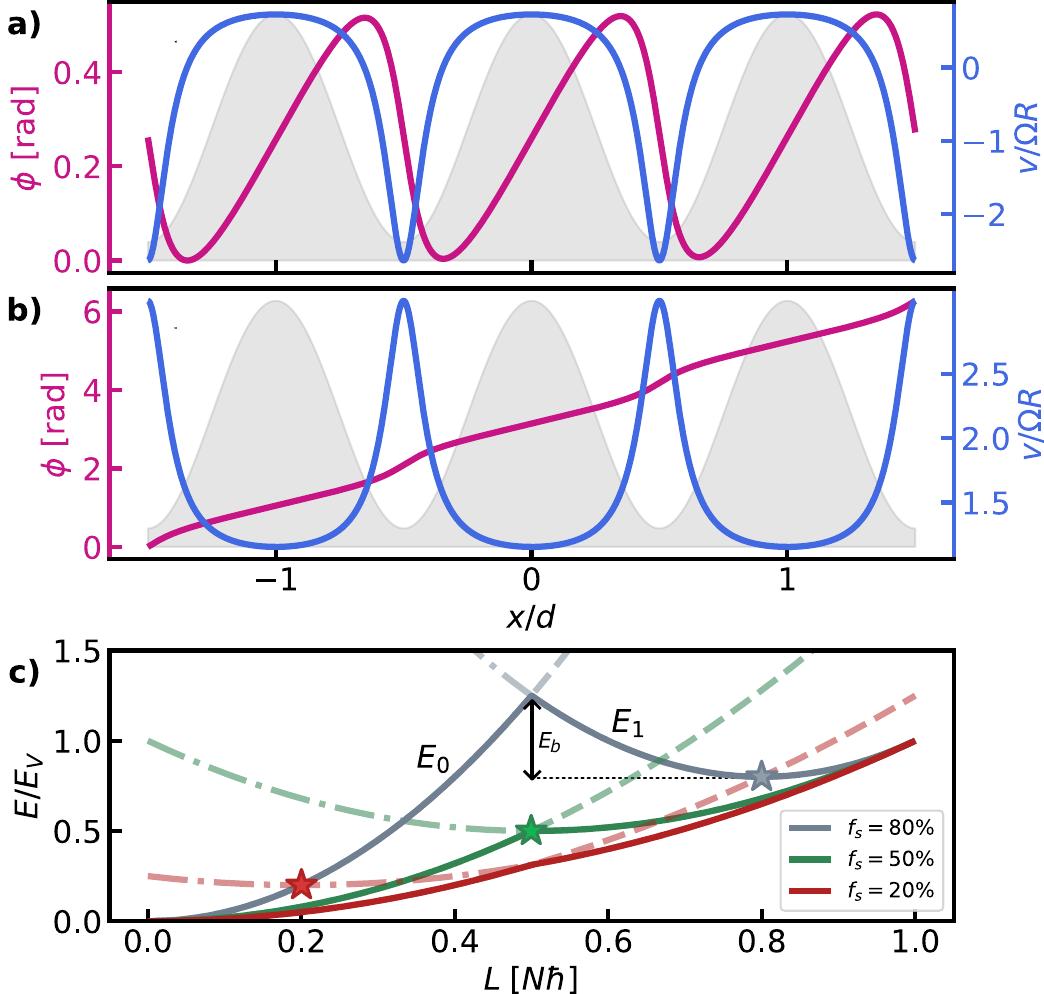}
    \caption{\textbf{Single fluid model.} Microscopic velocity field (blue) and corresponding phase profile (magenta) for a rotating supersolid with $f_s \approx$ 50\% in the classical manifold with zero circulation quanta (a), $w=0$, and with one circulation quantum (b), $w=1$. The supersolid density is plotted in gray. The horizontal axis is in units of the lattice period $d$. In this plot $\Omega/\Omega_c\approx 1.3$. c) Energy in the lab frame $E_0$ (dashed) and $E_1$ (dashed-dotted) versus the angular momentum $L$, for three different values of $f_s$. Thick lines indicate minimum energy states for a given $L$. $E_V=(N\hbar)^2/2I_c$ is the vortex energy ($\approx3\;\si{\micro\kelvin} \times k_B$ with our parameters). The stars indicate the local energy minima, corresponding to persistent currents.}
    \label{fig:PhaseFields}
\end{figure}

\textit{Introduction - }When the global gauge symmetry and translational symmetry are both spontaneously broken, the supersolid phase emerges, featuring a crystalline structure coexisting with superfluidity \cite{Gross, Chester, Le70}. After the original attempts in solid helium \cite{Balibar}, supersolids have been observed in quantum gases platforms such as Bose-Einstein condensates in cavities \cite{Esslinger}, with spin-orbit coupling \cite{SOC_Ketterle, Ch24} and with dipolar interactions \cite{PRLPisa, PRXInnsbruck, PRXStoccarda, 2DInnsbruck}. Fundamental properties assessed in these supersolid quantum gases could also play a role in more complicated condensed matter systems \cite{Ny17, Le19, Ch21, Sh20, Li23, Tr25, La24}.\\
The competition between superfluid flow and crystal rigidity requests modifications to the standard paradigm of superfluidity, some of which could be addressed in the inhomogeneous trapped systems available so far \cite{Science, Bi24, Ca24}. However, persistent currents in an annular geometry \cite{Stringari, leggett2006quantum}, one of the most intriguing phenomena related to superfluidity, are still experimentally unexplored. In superfluids, the hallmark of persistent currents is the quantization of the angular momentum per particle, $L/N = w\hbar$, with $w$ an integer \cite{Ry06}. The quantization arises from the single-valuedness of the macroscopic wavefunction, whose phase wraps linearly along the annulus, accumulating $w2\pi$ radians per turn. In contrast, supersolids host persistent currents with a reduced angular momentum per particle \cite{Te21}, $L/N = w f_s \hbar < w\hbar$, with $f_s$ the superfluid fraction of the supersolid \cite{Le70}. The phase wrapping along the ring still determines the quantization rule, but it coexists with a classical angular momentum sustained by the crystalline lattice. Partially quantized supercurrents pose new conceptual and experimental challenges. On the experimental side, usual measurements of annular supercurrents extract the winding number $w$ \cite{Mo12,Co14,Pa22}, which is not enough to demonstrate the partial quantization. On the conceptual side, the standard interpretation of the phenomenon is with a two-fluid model \cite{Te21}, where the supersolid consists of $Nf_s$ particles in a fully superfluid phase, and $N(1-f_s)$ particles behaving as a classical solid \cite{Si24, Za24, Ho21}. This mirrors the Landau two-fluid theory for a superfluid at finite temperature \cite{Landau41}. However, one notes that supersolids have a sub-unity superfluid fraction even at $T=0$ \cite{Bi24}, hence the normal component is not related to a thermal population, but rather to a 'crystalline component', which emerges from the same mean-field macroscopic wavefunction. Separating the system into two fluids with opposite behavior is therefore in contrast with the very nature of supersolids.\\
Here we solve this seeming contradiction, showing how rotating annular supersolids can be described with a single-fluid model, based on the continuity equation, in which the space-dependent phase profile of the unique wavefunction dictates the properties of the currents. We derive an analytical expression for the microscopic phase field, which allows us to design phase imprinting protocols to impose the desired velocity field into the supersolid ground state. We then introduce an innovative method to measure the angular momentum, designed to be sensitive to the partial quantization rule. We perform numerical simulations to benchmark these techniques in experimentally relevant conditions for the case of dipolar supersolids. Our model also applies to standard superfluids in which an external potential imposes a density modulation. We finally show that the standard two-fluid description emerges from our more fundamental single-fluid model in the proper limit, losing however the information on the phase.\\
\textit{Single-fluid model - }\label{sec:Model}
We consider a superfluid of $N$ particles with mass $m$, confined in a ring of radius $R$ and thickness $\delta \ll R$ so that the system can be considered one-dimensional. To account for a periodic modulation of the wavefunction, we introduce the superfluid fraction $0 \leq f_s \leq 1$, distinguishing supersolids from homogeneous superfluids ($f_s=1$) and crystals of independent droplets ($f_s=0$). 
It is directly connected to the depth of the density modulation by the upper bound derived by Leggett \cite{Le70}, which becomes equality in 1D systems \cite{Leggett1998}
\begin{equation}\label{eq:Leggett}
f_s =
\Big(\frac{1}{d}\int_{\text{unit cell}}\frac{\text{d}x}{\rho (x)/\bar{\rho}}\Big)^{-1},
\end{equation}
with $d$ the lattice constant, $\rho (x)$ the number density and $\bar{\rho} = N/2\pi R$ the average density, and $x \in [0, 2\pi R]$ the position along the 1D ring.\\
Our starting point is a single-component supersolid described by the density $\rho(x,t) = |\psi(x,t)|^2$, periodic in space, and velocity $v(x,t) = (\hbar/m)\nabla\phi (x,t)$, where $\psi$ is the complex many-body wavefunction and $\phi$ its phase.  
We consider rigid rotations at a constant velocity $V = \Omega R$, with $\Omega$ the angular velocity, so that the density satisfies $\rho(x,t) = \rho(x-Vt)$. In this case, the continuity equation  $\partial_t\rho+\partial_x (\rho v) = 0$ \cite{Ho21} becomes $V\partial_x\rho=\partial_x (\rho v)$, with $v(x)$ the microscopic velocity field. By integration, we obtain 
\begin{equation} \label{eq:velocityfield}
    v_w(x)=V(1-k_w/\rho(x)),
\end{equation}
where $k_w$ is an integration constant fixed by the irrotational condition $\oint v_w\dd x=w h/m$, and $w$ represents the winding number of the current (i.e. the phase accumulates a jump $\Delta \phi = \phi(2\pi R)-\phi(0)= w2\pi$) and $h$ the Planck constant. We get $k_w(\Omega)=\bar{\rho} f_s(1-2w\Omega_c/\Omega)$, 
were we have defined $\Omega_c=\hbar/2mR^2$.
The superfluid fraction $f_s$  coincides with the Leggett result in Eq. \ref{eq:Leggett}. A similar derivation of Eq. \ref{eq:Leggett} for $w=0$ can be found in \cite{Ch23}. \\
In the classical and superfluid limit ($f_s=0$ or $f_s=1$ respectively), all the atoms rotate with the same velocity. In the classical case, $v=V$, while for the superfluid the velocity is quantized, $v_w = w\hbar/mR$. 
In the supersolid phase, $0 < f_s <1 $, the velocity field is instead space-dependent, as reported in Fig. \ref{fig:PhaseFields}. 
Interestingly, even in the lowest manifold ($w = 0$), the velocity $v_0(x) = V(1-f_s\bar{\rho}/\rho(x))$ is different from zero, see Fig.~\ref{fig:PhaseFields}(a). The irrotational condition is satisfied thanks to two counterpropagating terms: the atoms in the density maxima ($v/V > 0$) move in the opposite direction of the superfluid background ($v/V < 0$). When the rotation is imparted by an external potential rotating at $V$, this can be seen as the density maxima being dragged by the external potential.

\begin{figure}
    \centering
    \includegraphics[width=1\columnwidth]{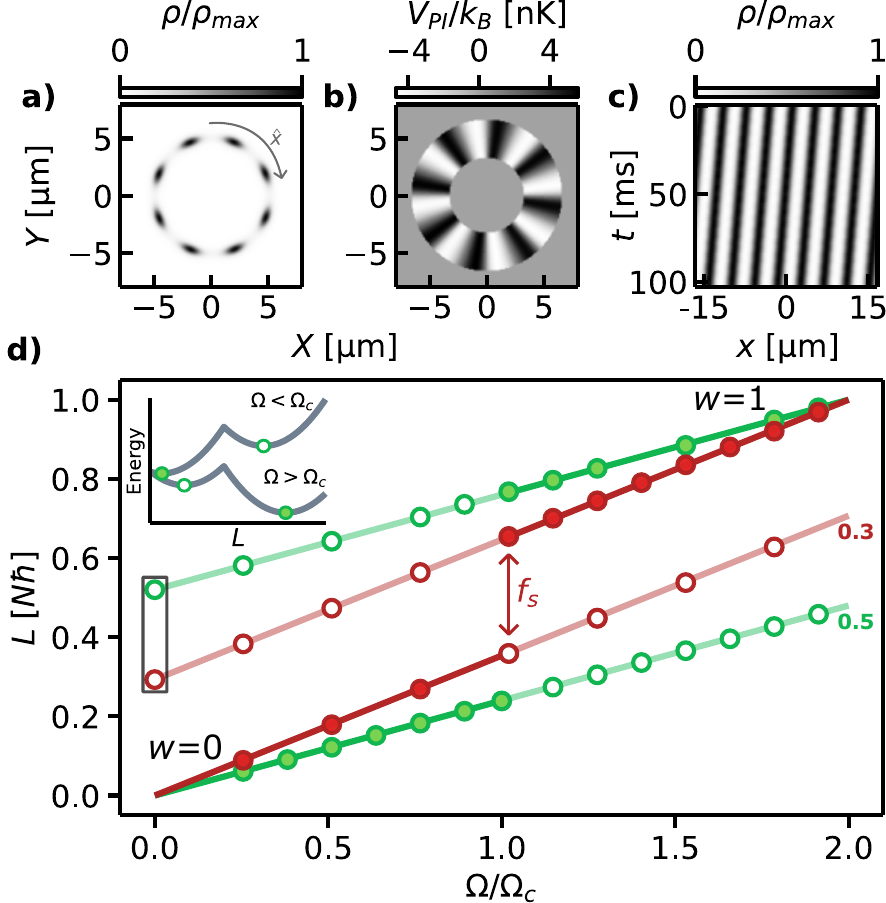}
    \caption{\textbf{Phase imprinting of angular momentum to the supersolid.} Supersolid ground state with $f_s \approx 0.5 $ (a) and corresponding 2D potential (b) used to phase imprint a current. c) Supersolid density $\rho(x,t)$ as a function of time, with $\Omega/\Omega_c\approx 0.7$ and $f_s\approx 0.5$, showcasing the rigid body motion of the density. d) Angular momentum $L$ vs angular velocity $\Omega$ for a rotating supersolid. The points are the results of numerical simulations, while the straight lines represent the prediction of the model, Eq. \ref{eq:L}. The corresponding superfluid fractions are indicated. Full points represent ground states in the corotating frame, while open points are metastable states, see inset. The points for $w=1$ and $\Omega = 0$, surrounded by the box, are persistent current states corresponding to the energy minima in Fig. \ref{fig:PhaseFields}(c).}
    \label{fig:Lvsomega}
\end{figure}

From the more general Eq. \ref{eq:velocityfield}, we derive the angular momentum of the supersolid also for $w>0$
\begin{equation}\label{eq:L}
    L =I_c(1-f_s)\Omega+Nwf_s\hbar,
\end{equation}
as well as its kinetic energy, $E_w=\int \dd x\rho v_w^2 /2$,

\begin{equation}\label{eq:energy}
E_w = \frac{(L-Nwf_s\hbar)^2}{2I_c(1-f_s)} + \frac{(f_sNw\hbar)^2}{2f_sI_c},
\end{equation}
with $I_c = NmR^2$ the classical moment of inertia. These results coincide with the two-fluid model in \cite{Te21} and contain indeed a classical contribution $L_c = I_c(1-f_s)\Omega$, increasing linearly with $\Omega$, and a superfluid contribution $L_s = Nwf_s\hbar$, quantized in units of $f_s\hbar$. \\
Depending on the value of $L$, the rotational states with minimum energy belong to different manifolds. In Fig. \ref{fig:PhaseFields}(c), we plot $E_0$ and $E_1$ for different $f_s$, highlighting the local energy minima of $E_1$. These are persistent current states, protected from decaying into $L=0$ by an energy barrier $E_b$. In the minima, $\partial_{L} E = \Omega =0$, thus the overall density doesn't move and the angular momentum is fully superfluid,  $L=L_s$. The barrier protecting the persistent current is formed by states with an additional classical component, causing the density to rotate in the opposite direction of the superfluid current (i.e. with $\Omega<0$ in Eq. \ref{eq:L}). These states have a smaller total $L$ than the persistent current, but a higher energy. The energy barrier reads $E_b=[(N\hbar)^2/(8I_c)][(1-2f_s)^2/(1-f_s)]\approx 1\;\si{\micro\kelvin}\times k_B$ for $f_s=0.8$ with our parameters (gray line in Fig. \ref{fig:PhaseFields}c). It disappears when $f_s \rightarrow 0.5$ and the local minimum of $E_1$ lies on the $E_0$ curve (green line). When $f_s<0.5$, the persistent current always has higher energy than the corresponding classical state with the same $L$ (red line).\\
To establish a connection with the two-fluid model, we average the microscopic current density $j(x) = \rho(x)v(x)$ over many unit cells, $j_{cg} = \int j(x) \text{d}x /(2\pi R)$. This operation corresponds to a coarse-grained description of the macroscopic current state.
We get $ j_{cg}=\rho_c v_c+\rho_s v_s$, where we have defined the crystal density $\rho_c=(1-f_s)\bar{\rho}$ and velocity $v_c=\Omega R$, and the superfluid density $\rho_s=f_s\rho$ and velocity $v_s=n\hbar /mR$. The current $j_{cg}$ is the same as for a superfluid at temperature $T>0$, once we replace the crystal with the thermal density \cite{Stringari}. \\
In this thermal case, the angular momentum also takes the same form as Eq. \ref{eq:L}, since the thermal component behaves classically. There is however a crucial difference between our model and the thermal two-fluid model. While in the latter the classical component is thermal and has no phase coherence, in our model the classical component is instead part of the same quantum many-body ground state. As such, we can derive a unique phase profile describing both the classical and superfluid motion, which would be impossible instead in the thermal case. The phase corresponding to a current with $w$ circulation quanta is $\phi_w = (m/\hbar) \int v_w(x) \text{d}x$, giving
\begin{equation} \label{eq:phix}
   \phi_w(x, \Omega)=\frac{m\Omega R}{\hbar}\left( x - \int_0^x \frac{\bar{\rho}f_s(1-2w\Omega_c/\Omega) \dd x'}{\rho(x')} \right),
\end{equation}
which we plot in Fig. \ref{fig:PhaseFields} together with the velocity field in Eq. \ref{eq:velocityfield}, for a given $f_s$.\\
\textit{Phase imprinting protocol. -} We now show how to rotate supersolids in annular potentials. A common way to inject angular momentum is to bring the system into equilibrium with a rotating container, as in the 'rotating bucket' experiment of superfluid helium \cite{Hess_Fairbank}. The analog for quantum gases consists of rotating the harmonic trap \cite{Ch00}, stirring the superfluid with a laser \cite{Ma00} or, as recently developed for dipolar systems, with a rotating magnetic field \cite{Kl22}. It is natural in this case to look at the angular momentum acquired versus the angular velocity $\Omega$ of the container. The knowledge of the phase profile in Eq. \ref{eq:phix}, instead, allows us to phase imprint the desired angular momentum to the supersolid ground state, through the application of a potential $V_{PI}(x) = \hbar \phi_w(x;\Omega)/\tau$ for a small time $\tau$. Experimentally, $V_{PI}(x)$ can be an optical potential engineered with a spatial light modulator. Phase imprinting techniques were employed in dipolar supersolids to excite Josephson oscillations \cite{Bi24}, in bosonic \cite{Ku18} and fermionic \cite{Pa22} superfluids to excite persistent currents in ring geometries, as well as in driven quantum gases with supersolid-like properties \cite{Li25}.\\
We start by simulating the ground state of $N=5\times 10^4$ atoms of $^{162}$Dy in a ring trap with radius $R=5\;\si{\micro\meter}$, solving the Gross-Pitaevskii equation (GPE), see \cite{SM}. Fig. \ref{fig:Lvsomega}(a) displays an example of a supersolid ground state, which we employ to compute $V_{PI}(x)$ from Eq. \ref{eq:phix}, shown in Fig.\ref{fig:Lvsomega}(b). After applying the potential $V_{PI}(x)$ for  $\tau\approx100\;\si{\micro\second}$, the supersolid rotates freely and we compute the acquired angular momentum $L = \int \psi^* \hat{L}_Z\psi \text{d}X \text{d}Y \text{d}Z$ \footnote{We call the spatial variable along the ring $x$, while the cartesian variable orthogonal to the $YZ$ plane is $X$.}. We repeat the phase imprinting protocol for different superfluid fractions, angular velocities, with and without one circulation quantum ($w= 0,1$). We find that the supersolid rotates rigidly for all the phase profiles we imprint, see Fig. \ref{fig:Lvsomega}(c), validating our initial hypothesis.
The results of the phase imprinting are shown in Fig. \ref{fig:Lvsomega}(d), where we plot $L$ versus $\Omega$. The GPE simulations confirm the predictions of the 1D model and the results of \cite{Te21}. We observe the classical angular momentum $L_{c} = (1-f_s)I_c\Omega$ of the $w=0$ manifold separated by a 'partially quantized' jump, equal to $f_sN\hbar$, from the $w=1$ manifold, as predicted by Eq. \ref{eq:L} (full circles). The jump happens at $\Omega = \Omega_c$. These states are the absolute minima of the co-rotating energy $E-\Omega L$, as sketched in the inset in Fig. \ref{fig:Lvsomega}, and would be populated in the hypothetical 'rotating bucket' experiment. With our protocol, instead, we can also prepare metastable states (open circles in Fig.\ref{fig:Lvsomega}(d)), corresponding to local minima of $E-\Omega L$, either in the $w=0$ ($\Omega>\Omega_c$) or in the $w=1$ ($\Omega<\Omega_c$) manifolds. In particular, we can excite the persistent current states, having $w>0$, $\Omega = 0$ and $L=L_s= Nwf_s\hbar$ (enclosed in the rectangle in Fig.\ref{fig:Lvsomega}(d)). Since these states are not ground states in the rotating frame, contrary to homogeneous superfluids, they cannot be populated in a 'rotating bucket' experiment,  unless following a hysteric path by increasing and decreasing the rotational velocity \cite{Te21}. Additionally, the phase imprinting protocol has the advantage of leaving the density distribution almost unaffected, contrary to the stirring techniques.\\
We make two experimental remarks. First, to match the position of the supersolid clusters with the potential $V_{PI}(x)$, we find the ground state in the presence of a shallow pinning potential \cite{Si24}, see \cite{SM}. The same solution can be adopted in experiments. Second, for the $w=1$ manifold, the imprinted phase should be adjusted such that the $0-2\pi$ jump lies in one density minimum, to avoid density excitations \cite{Ku18}.\\
\textit{Measure of the angular momentum - }
\begin{figure}
    \centering
    \includegraphics[width=1\columnwidth]{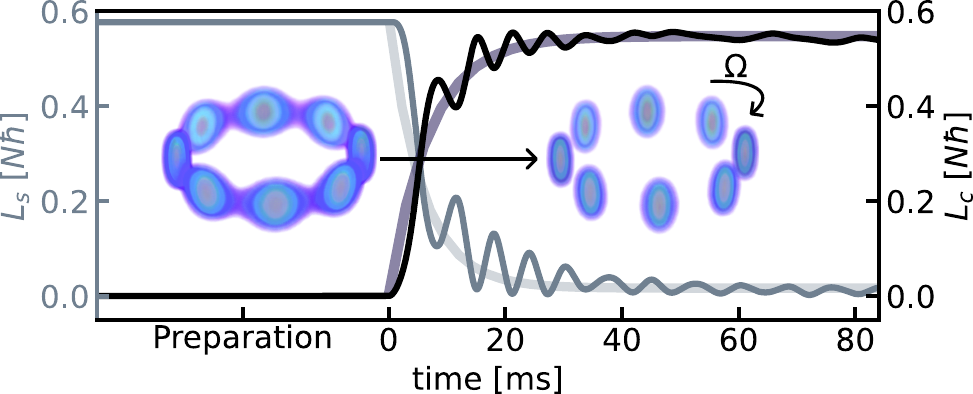}
    \caption{\textbf{Measure of the angular momentum of the supersolid.} Superfluid ($L_s$, gray) and classical ($L_c$, black) angular momentum as a function of time during a quench to the droplet crystal regime, $f_s\rightarrow 0$. The initial state is a persistent current with fully superfluid angular momentum, $L/N\hbar = f_s$. After the ramp, the angular momentum is transferred into the classical component, measurable through the classical rotational velocity of the droplets (inset).}
    \label{fig:angmom_meas}
\end{figure}
Many protocols have been developed to measure the angular momentum and its quantization in rotating superfluids  \cite{Ch00,Ku16, Mo12,Co14,Ec14, Fe25}. However, generalizing these protocols to supersolids is challenging \cite{Za24, Sc25}, see \cite{SM}. Here we propose an innovative protocol that takes advantage of the supersolid's mixed superfluid/classical rotational nature. We consider a supersolid with superfluid fraction $f_s$ rotating with some initial angular momentum $L_{\text{initial}} = L_c + L_s$, see Eq. \ref{eq:L}. We then ramp the system to the droplet crystal phase, where $f_s = 0$ and the angular momentum is purely classical, $L_{\text{final}} = I_c \Omega_{\text{final}}$. Due to the conservation of $L$, the superfluid angular momentum is transferred to the classical component. At this point, the rotational velocity of the isolated droplets is directly related to the initial angular momentum through $\Omega_{\text{final}} = L_{\text{initial}}/I_c$.
A measurement of the droplet crystal velocity through non-destructive in situ imaging, together with the knowledge of $I_c$, yields the angular momentum $L_{\text{initial}}$ of the initial current. Remarkably, this method is sensitive to the partial quantization, as it gives access to the full $L$ and not only to the winding number $w$ \cite{SM}. In Fig. \ref{fig:angmom_meas}, we simulate the measurement of a persistent current with $L_{\text{initial}}/N = f_s\hbar$, hence starting with $L=L_s$. The persistent current corresponds to $\Omega=0$, so the clusters don't move. We ramp down the scattering length $a_s$ to reach the droplet crystal phase. We plot $L_s/N = f_s (t)\hbar$ and $L_c/N = (1-f_s (t))I_c\Omega (t)$, with $f_s(t)$ and $\Omega(t)$ computed with Leggett's bound Eq. \ref{eq:Leggett} and the velocity of the density maxima, respectively. During the ramp, as $f_s$ gets lower, the velocity of the clusters increases, reflecting the transfer of angular momentum to the classical component. In the limit $f_s \rightarrow 0$, reached in our simulation after $t\approx50$ ms, the clusters' velocity absorbs the full initial angular momentum $L_{\text{initial}}$. A discrepancy of about $0.5\%$ is compatible with the residual $f_s$ at the end of the ramp. The superfluidity of the single droplets is instead negligible for our thin ring. We observe a weak excitation of an amplitude mode \cite{He24} corresponding to an oscillation of the clusters' population, which reflects in a corresponding oscillation of $f_s$ and clusters' velocity, visible in Fig. \ref{fig:angmom_meas}.
Experimentally, it is important to engineer an adiabatic ramp in the scattering length that doesn't excite additional modes, but at the same time doesn't require spending too much time in the droplet crystal phase, where the lifetime is strongly reduced \cite{PRLPisa}. Using the isotope $\mathrm{^{164}Dy}$, which exhibits a longer lifetime in the supersolid phase \cite{PRXInnsbruck}, could allow for longer ramps. We note that a similar mechanism of angular momentum transfer between classical and superfluid components was proposed to model glitches in neutron stars \cite{poli2023glitches}.\\
\textit{Partially quantized currents in optical lattices - } Similarly to supersolids, superfluids subjected to optical lattices exhibit a reduced superfluid fraction due to the explicit breaking of the translational symmetry \cite{Ch23}. We expect our microscopic model, designed for rigidly rotating supersolids, to also be valid for an infinitely rigid lattice.
\begin{figure}
    \centering
\includegraphics[width=1\columnwidth]{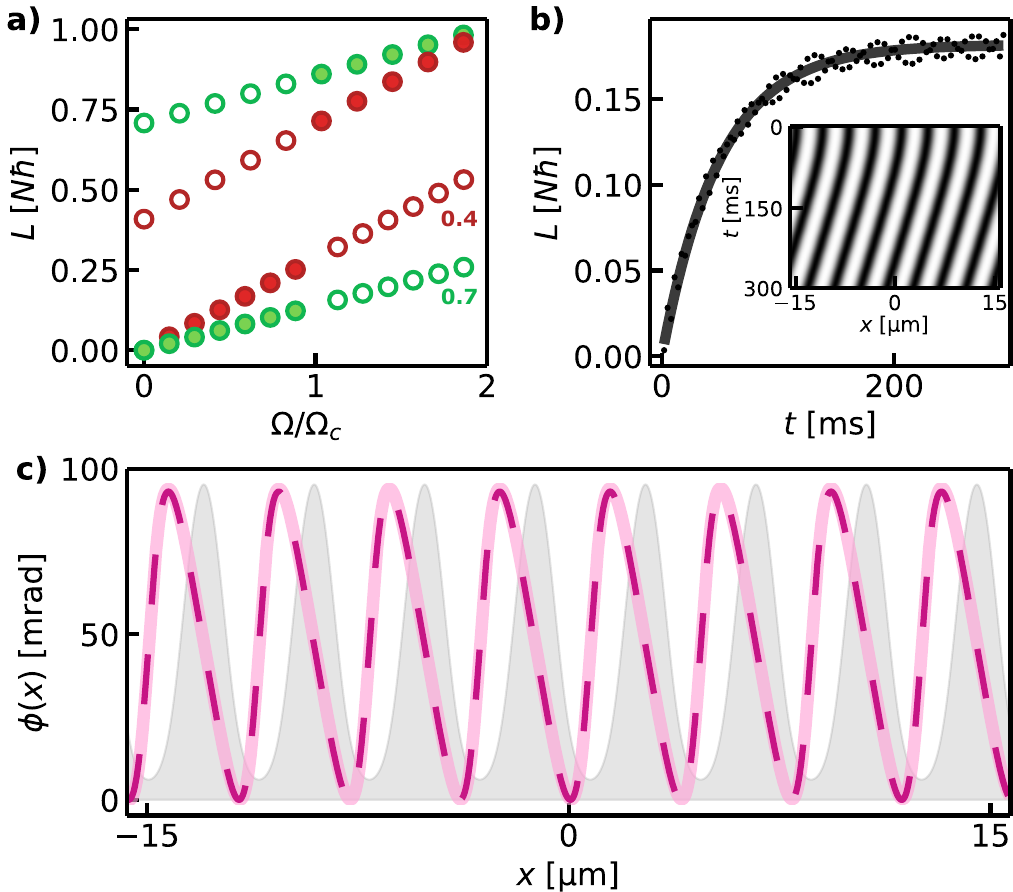}
\caption{\textbf{Superfluid in a rotating optical lattice.} a) Angular momentum $L$ vs angular velocity $\Omega$ for a superfluid in a rotating lattice for the $w=0$ and $w=1$ manifolds. Different colors correspond to different superfluid fractions, as displayed in the picture. b) Preparation of a classical rotational state ($w=0$) through a lattice rotation with angular velocity $\Omega(t) = \Omega(1-e^{-t/\tau})$, with $\Omega/\Omega_c\approx 0.6$, $\tau=50\;$ms. The points are the results of the simulation, while the continuous line is $L_c(t) = (1-f_s)I_c\Omega(t)$. The inset shows the density $\rho(x,t)$. c) Phase profile in the steady state of panel b) (lighter purple, thick) with the theoretical prediction based on the 1D model (purple, dashed). The density is plotted in gray for reference.}
    \label{fig:Lattice}
\end{figure}
To simulate this scenario, we again consider $N=5\times 10^4$ atoms of $^{162}$Dy atoms in a ring, but this time we keep the system far from the supersolid phase transition, see \cite{SM}. The ground state density in the ring potential is then uniform. To induce an artificially reduced superfluid fraction, we apply a lattice potential
with 8 lattice sites along the ring, emulating the supersolids in Fig. \ref{fig:Lvsomega}. 
We tune the superfluid fraction $f_s$ with the lattice depth (see \cite{SM}). The acceleration of the external lattice naturally constitutes an additional tool to impart classical angular momentum, as shown in Fig.\ref{fig:Lattice}(b) using an exponential ramp. To excite a generic state in the $w=1$ manifold, instead, we first inject $L_s$ through phase imprinting and then $L_c$ through a rotation of the lattice with the desired velocity $\Omega$. In Fig.\ref{fig:Lattice}(a), we summarize the measurements of $L$ as a function of the lattice final velocity $\Omega$, for $w=0$ and $w=1$. The results agree with the 1D model and the supersolid case, showing that similarly quantized supercurrents can also be explored in superfluids in optical lattices. We also evaluate the phase acquired by the superfluid during the lattice rotation for $w=0$, shown in Fig. \ref{fig:Lattice}(c). It perfectly matches the prediction of Eq. \ref{eq:phix}, confirming that the model correctly describes the rotational states of a modulated superfluid, no matter the origin of the modulation.\\
\textit{Conclusions - }We have shown how a single-fluid model correctly describes the dynamics of supersolids rotating in an annular geometry. The knowledge of the microscopic phase profile allows devising experimental protocols to excite and detect persistent currents with reduced angular momentum. Future developments in annular geometries will address the superflow decay mechanisms beyond the paradigm of homogeneous systems \cite{Mo12, Wr13, Pa22}. The tools we developed will be valuable in this context; for instance, to excite hypersonic currents or to monitor the time evolution of the angular momentum. Extensions to two-dimensional supersolids, where vortices have been excited via magnetostirring \cite{Ca24}, could further reveal the role of partial quantization, whose experimental evidence remains elusive \cite{Si22, Gallemi_Vortices,An21}. More broadly, our results may find application within the wider class of spatially modulated superfluids, including annular Josephson junctions \cite{Pe24}, vortex necklaces \cite{Ri25} and other atomtronic configurations \cite{Am22}.
\\

\begin{acknowledgments}
We thank M. Modugno and A. Ala\~na for providing a code to solve the eGPE equation. We thank T. Dardier for assisting with the simulations in the preliminary phase of the project. Useful discussions with S. Stringari, A. Recati and the Trento group and with G. Roati and his group are also acknowledged.
This work was funded by the European Union (European Research Council, SUPERSOLIDS, Grant No. 101055319). We acknowledge support from the European Union, NextGenerationEU, for the “Integrated Infrastructure initiative in Photonics and Quantum Sciences” (I-PHOQS Grant No. IR0000016, ID Grant No. D2B8D520, and CUP Grant No. B53C22001750006) and for PNRR MUR Project NQSTI-PNNR-PE4 No. PE0000023. G.B. acknowledges support from the European Union’s Horizon Europe research and innovation programme under the Marie Skłodowska-Curie grant agreement No. 101204101.
\end{acknowledgments}

\begin{center}
    \textbf{DATA AVAILABILITY}
\end{center}
The data that support the findings of this article are openly available \cite{data}.

\nocite{PRX, he2025exploring}
%

\newpage
\clearpage
\section*{Supplementary information}
\subsection{Numerical simulations}
We simulate the system dynamics by numerically integrating the extended Gross-Pitaevskii equation (eGPE):
\begin{equation} \label{eq:egpe}
\begin{split}
    i\hbar \partial_t \psi(\bm{r},t)=&\bigg[-\frac{\hbar^2\nabla^2}{2m}+V_{trap}(\bm{r})+g\abs{\psi(\bm{r},t)}^2\\&+\int \dd\bm{r}' V_{dd}(\bm{r}-\bm{r}')\abs{\psi(\bm{r},t)}^2\\
    &+ \gamma(\epsilon_{dd})\abs{\psi(\bm{r},t)}^3\bigg]\psi(\bm{r},t)
\end{split}
\end{equation}
where $g=4\pi\hbar^2a_s/m$ is the contact interaction parameter, $V_{dd}(\bm{r})=\frac{C_{dd}}{4\pi}\frac{1-3\cos^2\varphi}{r^3}$ is the dipolar interaction, with $\varphi$ being the angle between $\bm{r}$ and the dipoles direction $\hat{Z}$ and $C_{dd}=\mu_0\mu^2$ where $\mu_0$ is the Bohr magneton and $\mu$ is the modulus of the dipole moment. The external potential
\begin{equation}
V_{trap}(r, Z)=-V_0 e^{-(r-R)^2/2\sigma^2}+\frac{1}{2}m(\omega_r^2 r^2+\omega_Z^2Z^2)
\end{equation}
represents the sum of an harmonic potential with trapping frequencies $(\omega_r, \omega_Z)=2\pi\times (20, 100)~\text{Hz}$ and a ring potential with $R=5~\si{\micro\meter}$ and $\sigma=1~\si{\micro\meter}$ is chosen so that the phase transition between superfluid and supersolid is continuous in the ring with the chosen atom number. In this equation $r=\sqrt{X^2+Y^2}$.
The last term of Eq. \ref{eq:egpe} corresponds to the Lee-Huang-Yang quantum fluctuations contribution, where $\gamma(\epsilon_{dd})=\frac{128\hbar^2}{3m}\sqrt{\pi a_s^5}\text{Re}[\mathcal{Q}_5(\epsilon_{dd})]$,$\mathcal{Q}_n(x)=\int_0^1 \dd u(1-x+3xu^2)^{n/2}$ and $\epsilon_{dd}=C_{dd}/3g$ represents the ratio between dipolar and contact interactions strength. 
For details on the energy minimization methods employed to find the ground state in static simulation and to explore the system time evolution, see \cite{PRX}. 
The computations are performed on a box of $d_X\times d_Y\times d_Z = 24\times 24\times 24\; \si{\micro\meter}^3$ with a grid of $n_X \times n_Y \times n_Z = 256\times 256\times 64$ points.

\subsubsection{Supersolids}
\begin{figure}
    \centering
    \includegraphics[width=1\columnwidth]{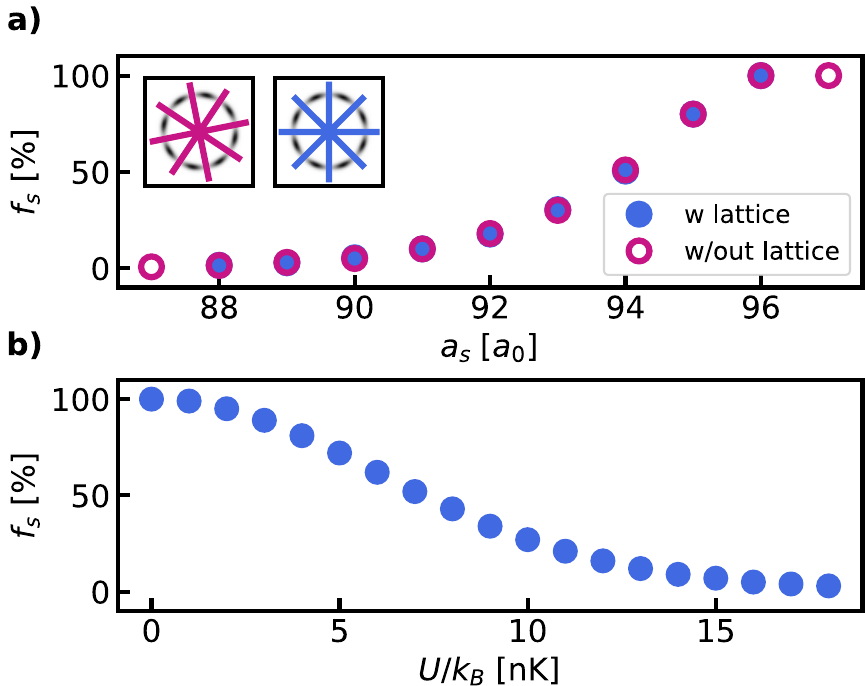}
    \caption{\textbf{Superfluid fraction and pinning lattice} a) Superfluid fraction of the ground state of the supersolid as a function of the $s$-wave scattering length $a_s$, in units of the Bohr radius $a_0$. The full blue circles are results of simulations in which a pinning lattice is added in the ground state while the hollow purple points are without the lattice. The inset shows two degenerate ground states differing by a rotation of the clusters. The two wheels are references for the eyes. b) Superfluid fraction $f_s$ of a superfluid in an external optical lattice, as a function of the lattice depth $U$.}
    \label{fig:pinning}
\end{figure}

The ground state of the stationary form of Eq. \ref{eq:egpe} is a supersolid if the $s$-wave scattering length $a_s$ is set below a critical threshold, which depends on the chosen atom number and the shape of the external trap. With our parameters, the ground state has 8 clusters arranged along the ring, see Fig. \ref{fig:Lvsomega}, and the transition from the superfluid to the supersolid happens around $a_s \approx 96\;a_0$, see Fig. \ref{fig:pinning}. 
Since the energy functional in Eq. \ref{eq:egpe} is rotationally invariant in the azimuthal direction ($x$ in the main text), infinite degenerate ground states exist, differing only in the positions of the clusters along the ring.
To remove this degeneracy and select a well defined ground state within different numerical realizations, we add a shallow lattice along the ring to pin the clusters.
This is also crucial to minimize the mismatch between the density modulation and the phase profile imprinted, as described in the main text. 
The pinning lattice is $V_{pin}(x)=U\sin[q x]$ with $q=8$ and amplitude $U$ chosen small enough (about 1 nK) to remove the degeneracy in the azimuthal direction without changing the superfluid fraction of the supersolid, see Fig.\ref{fig:pinning}(a). Immediately after the application of the phase imprinting potential, we remove the pinning lattice, which is off in the following dynamics.

\subsubsection{Modulated superfluid}
For the simulations of Fig. \ref{fig:Lattice} in the main text, we fix $a_s=110\;a_0$ to keep the system superfluid, far from the supersolid phase transition.
To synthetically lower the superfluid fraction in this case, we increase the amplitude of the pinning lattice. 
We keep $q=8$ to have the same number of clusters as in the supersolid case.
We tune the superfluid fraction by changing the amplitude of the lattice, see Fig. \ref{fig:pinning}(b).
By making the lattice rotate, substituting $V_{pin}(x)\rightarrow V_{pin}(x-\Omega(t)t)$, it is then possible to give classical angular momentum to the system. We ramp up the angular velocity $\Omega$ by accelerating the lattice slowly enough, so that the density evolves rigidly as $\rho(\theta-\Omega(t)t)$. 
It is convenient to introduce the effective angular velocity
\begin{equation}
    \Omega_{eff}(t)=\frac{\dd\left[\Omega(t)t\right]}{\dd t}=\Omega(t)+t\frac{\dd\Omega(t)}{\dd t}\,\,.
    \label{eq:omega_eff}
\end{equation} 
Ramping $\Omega$ with constant acceleration up to a final value $\Omega_0$, reached after a time $\tau$, we have $\Omega(t)=\Omega_0 t/\tau$ and $\Omega_{eff}=2\Omega_0 t/\tau$.
In this case, the angular momentum increases linearly with time. 
Evaluating $L$ during the dynamics, we can map $L(t)\to L(\Omega)$, and build the $w=0$ manifold of Fig.\ref{fig:Lattice}(a). 
For the $w=1$ manifold, we first imprint the phase $\phi_w(x,\Omega=0)$ and then accelerate the lattice following the same procedure. 
Note that, in this case, the phase imprinting can be used because the imprinted state has no classical rotational component ($\Omega=0$). Before accelerating the lattice, the density does not move, and the external potential doesn't compete with the current.
To reach a steady rotational state with a fixed $L$, we accelerate the lattice to have $\Omega_{eff}(t)=\Omega_0(1-e^{-t/\tau})$.
Solving Eq.\ref{eq:omega_eff} for $\Omega(t)$ gives $\Omega(t)=\Omega_0\left[1+\frac{\tau}{t}e^{-t/\tau}\right]+\frac{c}{\tau}$, where $c$ is an integration constant determined by the condition $\Omega(t_0)=0$. 
The results for this procedure are shown in Fig. \ref{fig:Lattice}(b).

\subsection{Time of flight measurements of the angular momentum}

The protocol we propose in the main text to measure the angular momentum relies on the possibility to tune the superfluid fraction of the supersolid. For homogeneous superfluids, many other methods are known. For example, the presence of a persistent current lifts the degeneracy of elementary excitations due to the Doppler effect, and the energy splitting is proportional to the angular momentum \cite{Ch00,Ku16}. Another class of measurements relies instead on the interference pattern of the system after a free expansion. If the expanding gas is made to interfere with a BEC at rest, the appearance of a typical spiral pattern allows counting the winding number \cite{Co14, Ec14} and also reconstructing the local phase around the ring. However, the generalization of these protocols to supersolids is difficult or even impossible \cite{Za24, Sc25}. For example, the splitting of the two sound modes due to the Doppler effect in a supersolid \cite{Za24} gives no direct information on the angular momentum, contrary to the BEC case. Moreover, although persistent currents affect the distribution in time of flight also in supersolids \cite{Sc25}, the interference of the coherent density clusters of the supersolid dominates. Whether the partial quantization of the angular momentum can be resolved in this experiment is unclear.
\begin{figure}[t]
    \centering
    \includegraphics[width=1\columnwidth]{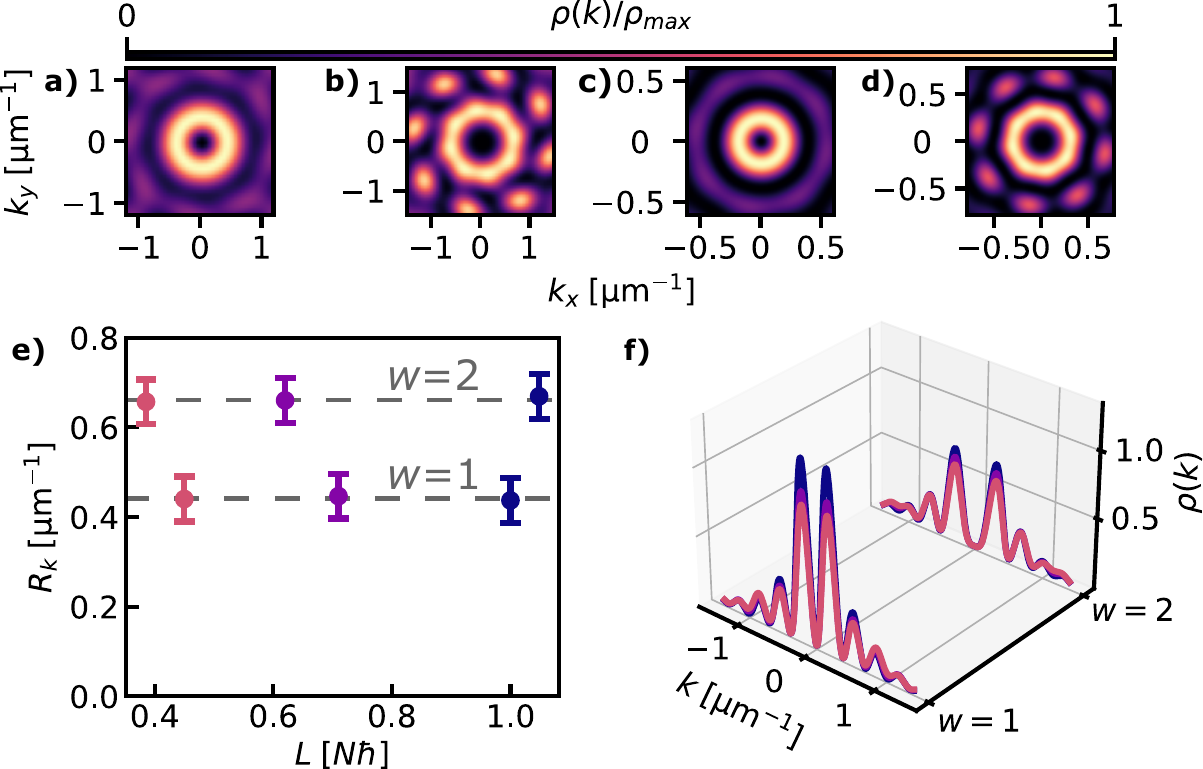}
    \caption{\textbf{Expansion of a rotating supersolid} 
    (a-b) Density distributions after the expansion in the harmonic trap, after removal of the ring potential, for supersolids with (a) $w=1$ and (b) $w=2$. (c-d) Momentum distribution obtained via Fourier transform of the in-situ wave-function for supersolids with (c) $w=1$ and (b) $w=2$. (e) Size of the hole in the density after expansion, $R_k$, as a function of the angular momentum $L$ of the rotational states. The errorbars correspond to the uncertainty due to the numerical grid. The horizontal lines are fits that show no dependence on the angular momentum for fixed $w$. (f) One-dimensional cut along $k_y=0$ of the Fourier transforms, for different $w$. Each plot has a different $L$, corresponding to the points in (e).}
    \label{fig:tof}
\end{figure}
\\
Here we show simulations of the expansion of supersolids hosting a persistent current, without a BEC reference at the center. For rotating homogeneous superfluids, the momentum distribution displays a hole at the center $k=0$, due to the centrifugal force, with a radius $R_k$ quantized accordingly to the winding number of the current \cite{Mo12}. In our simulations, we initialize currents in the supersolid with different $f_s$, $L$, and $w$. To avoid the computational cost of a full time of flight, we perform a momentum focusing technique, i.e. we release the atoms from the ring and let them expand in the remaining harmonic trap \cite{he2025exploring}. At the time of the release, we change the scattering length to the fixed value $110$ $a_0$ to minimize and equalize dipolar effects in the expansion. Figs.\ref{fig:tof}(a-b) show examples of the density after the expansion, for $w=1,2$. We checked that our results give good qualitative information on the momentum distribution by comparing the obtained densities with the analytic Fourier transform of the in situ density, see Figs.\ref{fig:tof}(c-d). A hole is visible for both $w=1,2$, and its size $R_k$ grows with $w$. In Fig.\ref{fig:tof}(e) we show $R_k$ as a function of the angular momentum of the system, for different winding numbers. Interestingly, $R_k$ depends strongly on the winding number, but is insensitive to the partial quantization of $L$. This result can be understood from the expression of the phase $\phi$ of a rotating supersolid, in Eq. \ref{eq:phix}. Through some algebra, we can recast $\phi$ in the following form
\begin{equation}
    \phi_w(x,\Omega)=w \frac{x}{R}-(2w\Omega_c -\Omega)\frac{mR}{\hbar}\left(x-\int_0^x \frac{\bar{\rho}f_s}{\rho(x')}\dd x'\right)\, .
    \label{eq:phase_w}
\end{equation}
The first term is equal to the phase of a regular superfluid with winding number $w$, and it fixes the quantization of $\Delta \phi = w2\pi$ around a loop in the ring. 
The second term, proportional to $\phi_0(x,2w\Omega_c-\Omega)$, is instead peculiar to the supersolid phase and is responsible for the partial quantization of $L$.
Evaluating the momentum distribution as $\rho(k)\sim\abs{\int \dd x\; e^{i(kx+\phi(x))}}^2$ we find that only the first term affects the $k\approx0$ components, where the hole forms.
The hole size is then only dependent on the winding number through the first term of Eq.\ref{eq:phase_w}.
This explains why $R_k$ is not a suitable observable to measure the angular momentum of a supersolid.

\end{document}